\documentclass[journal=jmcmar,layout=twocolumn]{achemso}

\usepackage[font=small]{caption}

\author{Michal H. Kol\'{a}\v{r}}
\affiliation[IOCB CAS CR]
{Institute of Organic Chemistry and Biochemistry of the Czech
Academy of Sciences, Flemingovo nam. 2, 16610 Prague, Czech
Republic}
\email{michal.kolar@uochb.cas.cz}
\alsoaffiliation{Computational Biophysics, German Research School 
for Simulation Sciences GmbH, 52425 J\"ulich, Germany}
\alsoaffiliation{Institute for Advanced Simulations (IAS-5), 
Forschungszentrum Jülich GmbH, D-52425 J\"ulich, Germany}
\author{Paolo Carloni}
\affiliation{Computational Biophysics, German Research School 
for Simulation Sciences GmbH, 52425 J\"ulich, Germany}
\alsoaffiliation{Institute for Advanced Simulations (IAS-5), 
Forschungszentrum Jülich GmbH, D-52425 J\"ulich, Germany}
\alsoaffiliation{Institute of Neuroscience and Medicine (INM-9),
Forschungszentrum Jülich GmbH, D-52425 J\"ulich, Germany}
\author{Pavel Hobza}
\affiliation[IOCB CAS CR]
{Institute of Organic Chemistry and Biochemistry of the Czech
Academy of Sciences, Flemingovo nam. 2, 16610 Prague, Czech
Republic}
\alsoaffiliation{Regional Centre of Advanced Technologies 
and Materials, Department of Physical Chemistry, Palack\'{y}
University, Olomouc, 771 46 Olomouc, Czech Republic}

\title{Statistical Analysis of $\sigma$-Holes: A Novel 
Complementary View on Halogen Bonding}

\begin{document}



\begin{abstract}
To contribute to the understanding of noncovalent binding 
of halogenated molecules with a biological activity, electrostatic 
potential (ESP) maps of more than 2,500 compounds were thoroughly 
analysed. A peculiar region of positive ESP, called $\sigma$-hole,
is a concept of central importance for halogen bonding. We aim at
simplifying the view on $\sigma$-holes and provide general trends in 
organic drug-like molecules. The results are in fair agreement
with crystallographic surveys of small molecules as well as of 
biomolecular complexes and attempt to improve the intuition of
chemists when dealing with halogenated compounds.
\end{abstract}


\section{Main Text}

Halogen bonding (XB) is an attractive interaction between a halogen 
atom and a Lewis base. It has received appreciable attention in the
last decade. It has found a distinguished place in drug \cite{1,2,3}
and material design \cite{4}. The electrostatic and London 
dispersion interactions are the most prominent effects contributing 
to the stabilization of XB complexes \cite{5,6,7}.

The anisotropy of electrostatic potential (ESP) (i.e. the $\sigma$-hole)
has been recognised by quantum chemical calculations on halogen 
atoms \cite{8}, and later also on the atoms of IV, V and VI 
groups \cite{9,10}. The properties of halogen $\sigma$-holes have 
recently been abstracted into two characteristics: i) $\sigma$-hole 
magnitude, which stands for the value of the most positive electrostatic 
potential on the 0.001 e/bohr$^3$ electron density isosurface, and 
ii) $\sigma$-hole size, i.e. the spatial extent of the positive 
region \cite{11}. Although the $\sigma$-hole magnitude and size were 
initially introduced only for molecules with a $C_{2v}$ symmetry point 
group, it has been shown that such characteristics provide 
a simplified but relevant view on ESP anisotropy and XB. The magnitude 
correlates with the strength of XB \cite{6} while the size is related to the 
directionality of XB, understood as an angular channel, where the 
interaction between the molecules is attractive \cite{11}.

Here, we propose a generalization of the concept of $\sigma$-hole 
magnitude and size for non-symmetric organic molecules. Further, 
we apply such descriptors to characterize halogenated molecules 
in ZINC database \cite{12,13}. This database is a keystone of 
virtual screening \cite{14} which is a starting point for numerous 
drug-design studies. Thus, we aim at elucidating the typical $\sigma$-hole 
of a drug candidate. So far, we have calculated and analysed ESPs of 
331 molecules containing chlorine, 1,267 molecules containing bromine
and 1,003 molecules containing iodine. All of these molecules have at
least one C--X (X=Cl, Br, I) covalent bond. The actual number 
of $\sigma$-holes analysed was slightly higher, because some of 
the molecules comprised more than one halogen. From ZINC, we selected 
neutral purchasable molecules. The upper bound for molecular weight 
was 250, 300 and 350 Da for chlorinated, brominated and iodinated 
compounds, respectively.

All of the compounds were subjected to full energy minimization using 
a hybrid B3LYP functional with a 6-31G* basis set for all atoms except
for iodine, for which a LANL2DZ basis with a pseudopotential for the 
inner-core electrons was used. The default convergence criteria of the
Gaussian09 program package \cite{15} were adopted (maximum 
force $< 4.5\cdot 10^{-4}$ a.u., root-mean-square force 
(RMS) $< 3.0\cdot 10^{-4}$ a.u., maximum 
displacement $<1.8\cdot 10^{-3}$ a.u., RMS 
displacement $<1.2\cdot 10^{-3}$ a.u.). The electrostatic potential 
maps were calculated on a three dimensional grid from the electron 
densities at the B3LYP/def2-QZVP level.

The $\sigma$-hole magnitude was defined, like in Ref. \citenum{10}, 
as the value of the most positive (or the least negative) ESP lying 
on the halogen boundary, arbitrarily defined as the 0.001 e/bohr$^3$ 
electron density isosurface \cite{16}. Only the isosurface lying in 
a certain half-space was considered.

\begin{figure}[tbh]
\includegraphics{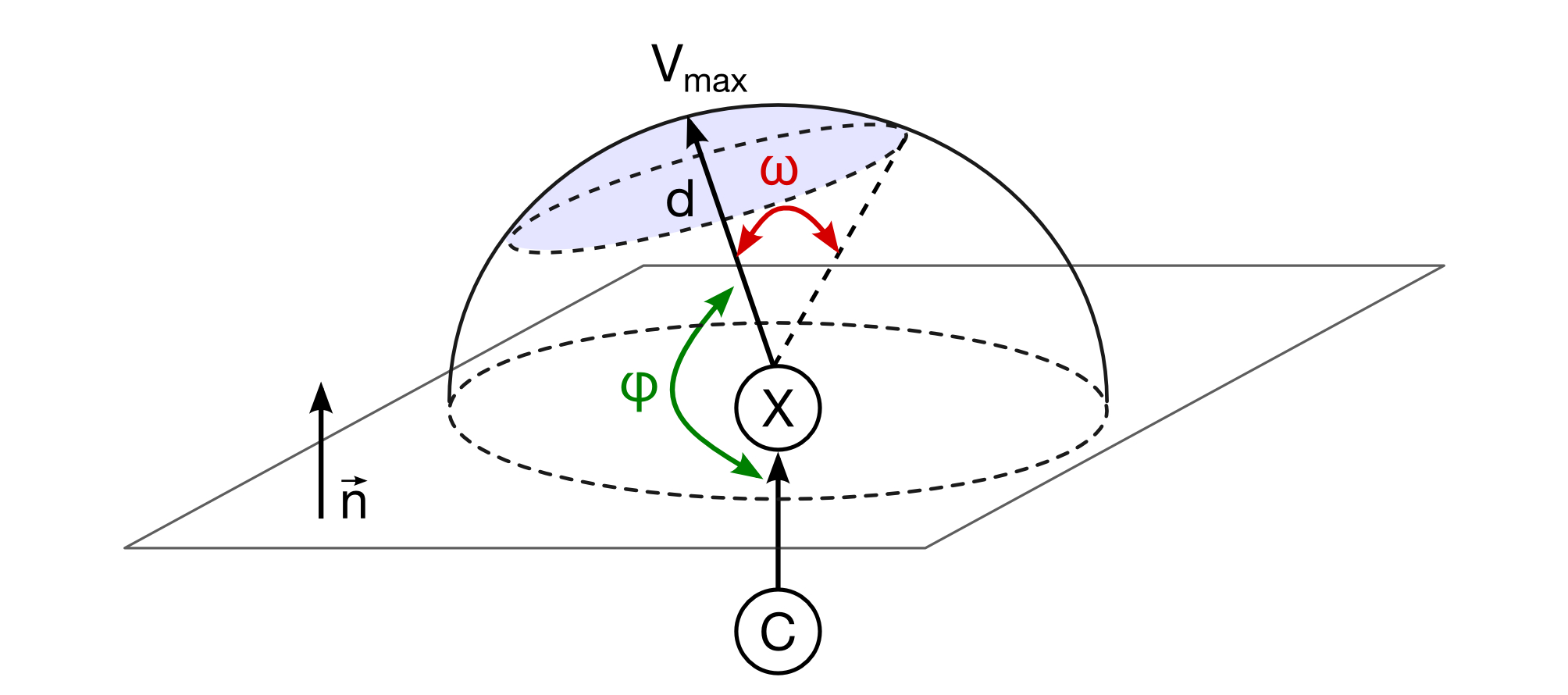}
\caption{The definitions of $\sigma$-hole characteristics.}
\label{fig:def}
\end{figure}

The half-space was defined by the position of the halogen and by 
a plane whose norm $\mathbf{n}$ was collinear with the vector of the halogen and 
its closest atom (carbon) (see Figure \ref{fig:def}). We further defined 
distance $d$ between the halogen and the point with the $V_{max}$, C--X--$V_{max}$
angular deviation $\phi$ and threshold angle $\omega$ between 
the $\sigma$-hole boundary and the vector $\mathbf{d}$ (Figure \ref{fig:def}).
Since other atoms may induce the most positive ESP with $\phi$ of 
about 90$^\circ$, the magnitude is found as the local maximum 
of the ESP closest to the elongation of the C--X bond.

\begin{figure}[tbh]
\includegraphics{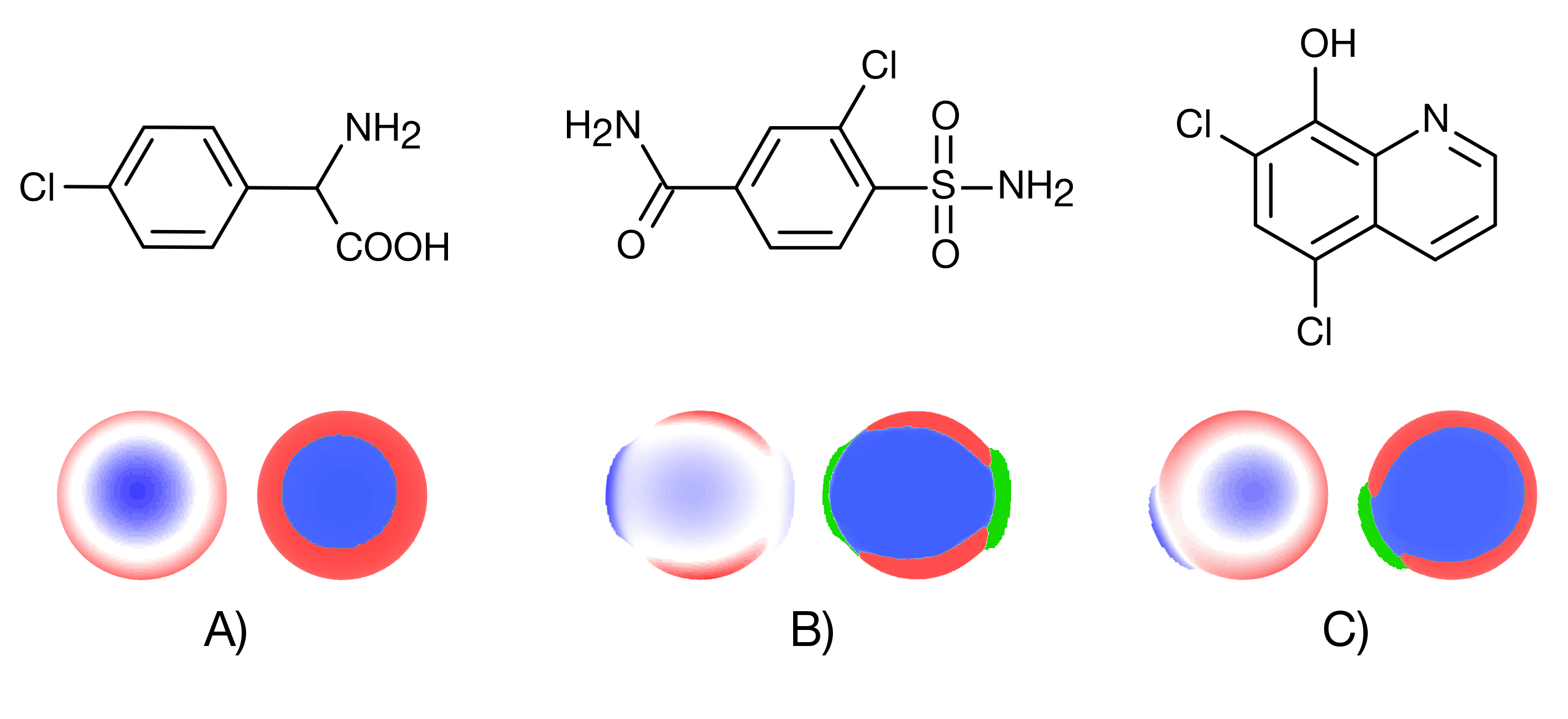}
\caption{Top views on three ESP maps of chlorine and their 
respective simplifications used for analysis. In the maps, 
the scale goes from the negative red trough white zero to the positive 
blue. In the simplifications, the $\sigma$-hole region is in blue, 
the negative ESP in red and the positive ESP excluded by clustering 
algorithm from the $\sigma$-hole region is in green;
A) Baclofen (ZINC00000061), B) 4-chloro-3-sulfamoylbenzamide 
(ZINC00002088), C) Chloroxine (ZINC00001131), 
(ESP of the chlorine in para- position wrt the hydroxyl).}
\label{fig:esps}
\end{figure}

The $\sigma$-hole size for non-symmetric molecules was defined as 
the area of the 0.001 e/bohr$^3$ electron density isosurface where
the ESP was positive. As shown in Figure \ref{fig:esps}, there may 
be molecules with a complicated shape of the positive ESP making 
the total area of positive ESP useless for $\sigma$-hole 
characterisation. Thus, we applied a clustering algorithm to select
only such a part of the positive surface that reflects the topology
of the ESP. It means that the area of positive ESP, which was induced
by the surrounding chemical entities, was discarded from the total
positive ESP area.

In order to obtain an approximately rounded boundary of the area 
of discrete set of positive ESP points, the size of the $\sigma$-hole 
was obtained as follows:

\begin{enumerate}
\item The point with the most positive ESP closest to the 
elongation of the C--X bond was localized.
\item For each grid point $P$ with a positive ESP, the angle 
$V_{max}$--X--P was calculated. A histogram of these angles 
was constructed.
\item We counted all the points up to the first local minimum 
of the histogram and multiplied the sum by the area per 
point. The minimum was denominated as threshold angle $\omega$.
\end{enumerate}

Such an approach selects only those positive ESP points, which
deviate from the ESP maximum by less than the value of threshold 
angle $\omega$.

Obviously by this approach, no $\sigma$-hole size could be assigned 
to halogens with overall negative ESP, even though such negative
regions have also been denominated as $\sigma$-holes before \cite{9}.
However here, only eight instances out of the entire set of 
drug-like organic compounds have been identified with no positive 
ESP region. Thus, a positive $\sigma$-hole is a general feature 
of such compounds.

\begin{figure}[tbh]
\includegraphics{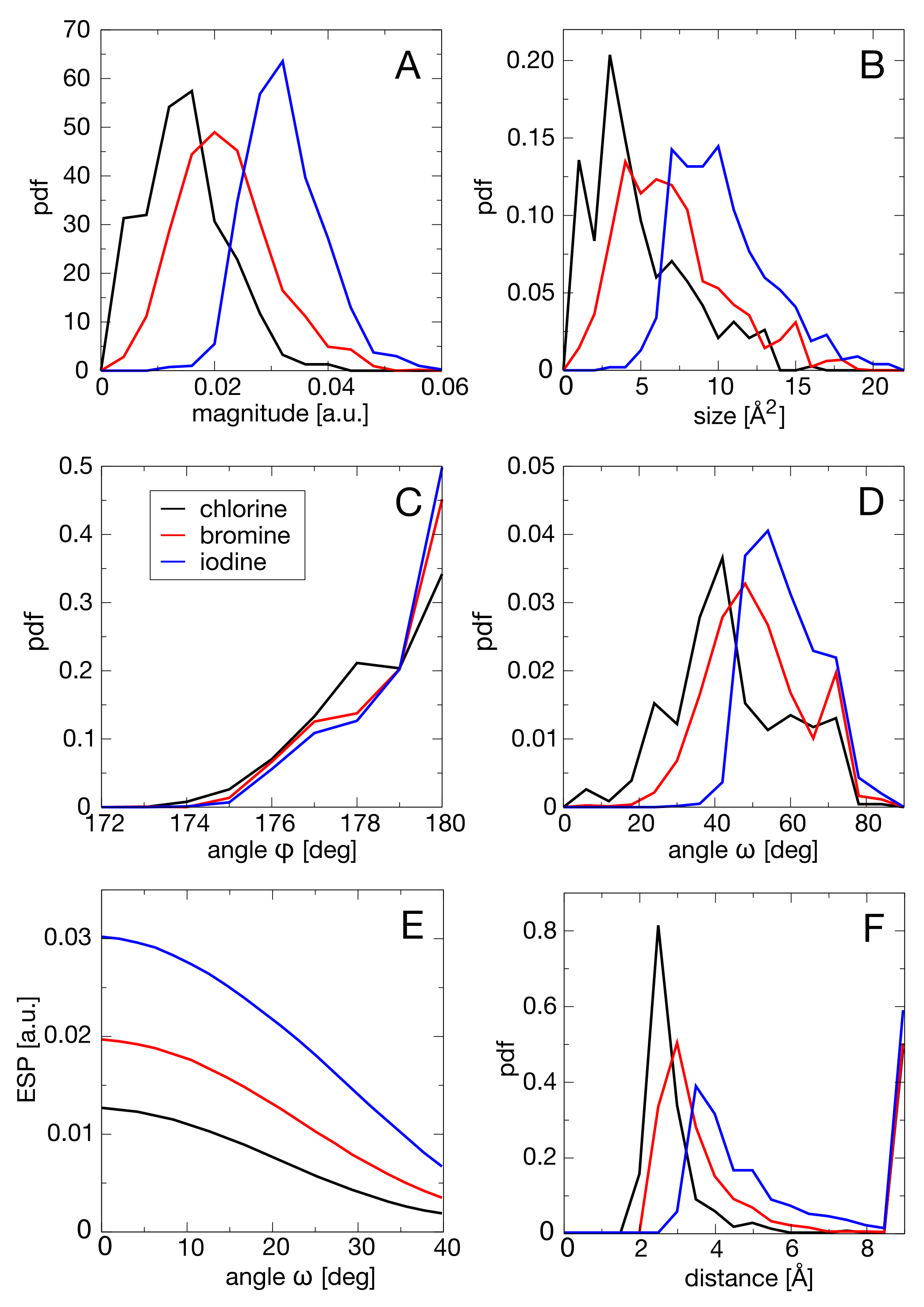}
\caption{Probability density functions (pdf) of A) the magnitude, 
B) the size, C) angular deviation $\phi$ D) threshold angle $\omega$.
E) The dependence of the ESP on angle $\omega$ and F) the pdf of 
the $\sigma$-hole range as the boundary of the positive and 
negative ESP in the X--$V_{max}$ direction. The pdf was 
summed for distances larger than 9.0 \AA.}
\label{fig:dists}
\end{figure}

The probability density functions (pdfs) are summarized 
in Figure \ref{fig:dists}. The magnitudes of the $\sigma$-holes 
(Figure \ref{fig:dists}A) show the previously known 
trend\cite{5,17,18} that the maximum of the ESP increases with 
the increasing atomic number of the halogen. The pdfs are 
gaussian-like with the mean values of 0.0127, 0.0198 and 
0.0304 a.u. for chlorinated, brominated and iodinated molecules. 
On average, the iodinated molecules thus have more than a twice
more positive $\sigma$-hole than the chlorinated ones.

The deviation angle $\phi$ (Figure \ref{fig:def}) is higher 
than 178$^\circ$ for about 50 \% of the set and higher 
than 172$^\circ$ for almost the entire set. Surprisingly enough,
the spatial position of $V_{max}$ in organic drug-like molecules 
is hence only little affected by the nature of the molecule or,
in other words, by the chemical environment of the halogen. It
implies that the electrostatic interaction of $\sigma$-holes supports 
the strictly linear arrangement of halogen bonds (see below). Finally,
the high values of angle $\phi$ provide the justification of placing 
a fixed positively charged particle on top of halogen atoms in 
molecular mechanistic studies \cite{19,20}.

The sizes of $\sigma$-holes (Figure \ref{fig:dists}B) increase with 
the atomic number of the halogen. Obviously, the size correlates 
well with the angle $\omega$ (R$^2$=0.87). Moreover with the increasing 
atomic number of the halogen, the $\omega$ pdfs shift to higher 
values and their width decreases. Consequently, the size variability 
of the chlorine $\sigma$-holes is higher than in the case of the 
iodine ones. The size correlates with the magnitude (R$^2$=0.91),
which was observed previously for symmetric molecules \cite{11}.

The ESP projections on the 0.001 e/bohr$^3$ isodensities provide 
information on the electrostatics in the vicinity of a compound.
We calculated the average distance of the $V_{max}$ point from 
the halogen, which was 1.93, 2.04 and 2.19 \AA ~for chlorinated, 
brominated and iodinated compounds, respectively. However, it is
often overlooked that the ESP changes with the distance from the
halogen and consequently that the $\sigma$-hole as a region of 
the positive ESP may vanish at longer distances. We calculated 
the range of the $\sigma$-hole as the distance from the halogen, 
where the ESP changes its sign from positive to negative. The direction 
of X--$V_{max}$ was evaluated (Figure \ref{fig:dists}F). The maxima 
of the pdfs of the $\sigma$-hole ranges were 2.5, 3.0 and 3.5 \AA from 
the chlorine, bromine and iodine, respectively, which is 1.25-, 1.50-
and 1.75-multiple of their vdW radii. Importantly, the $\sigma$-holes 
of more than 24 \% of chlorinated, 25 \% of brominated and 30 \% of 
iodinated compounds are preserved beyond 9 \AA ~from the halogen, 
which likely contributes to the middle-range stabilization of XB
complexes.

As stated in our previous work \cite{11}, ``the channel allowing 
the approach of hydrogen fluoride to halobenzenes is the narrowest
for Cl and the broadest for I,'' which is supported here by 
the pdfs of angle $\omega$ (Figure \ref{fig:dists}D). Thus the angular 
freedom of electron donors to establish an attractive interaction 
(i.e. with positive stabilization energy) with iodinated compounds
should be larger than with chlorinated or brominated ones. This is,
however, not reflected in crystallographic studies, from which it 
has resulted that the XBs involving iodine tend to be linear more
than those with chlorine and bromine \cite{21,22,23,24}. This merely 
agrees with the pdfs of angle $\phi$ (Figure \ref{fig:dists}C), 
although the differences are rather small. Furthermore, 
Figure \ref{fig:dists}E shows the angular profile of the ESP 
averaged over the set of molecules. The profile is the steepest for 
iodine compounds. Hence, the force acting on the electron donor 
is the largest when iodine is involved, favouring the linear 
arrangement of XB more than in the cases of bromine and chlorine. 
In other words, the electrostatic energy penalty needed to deform 
the linear arrangement is higher for iodine than for chlorine.

We have to distinguish between directionality, previously determined 
as a solid-angular channel allowing attractive interaction, and 
tendency to linearity. Directionality seems to be well captured 
only by the size of the $\sigma$-hole. This means that directionality 
increases on passing e.g. from chlorobenzene to iodobenzene. 
The tendency to linearity is expressed by the angular variations 
of electric field and thus it depends on both the magnitude and 
size of the $\sigma$-hole. Both, the magnitude and the size of 
the $\sigma$-hole, increase with the atomic number of halogen, 
but the increase of the magnitude is steeper.

In biomolecular complexes, it was observed \cite{24} that the angular 
variations of C--X$\cdots$O are significantly lower for iodine when 
compared to bromine and chlorine. This agrees well with the 
statistical analysis of $\sigma$-hole sizes and angles $\omega$ while 
it remains slightly puzzling for halogen with such large polarisability.

\section{Conclusions}

To summarize, we have presented an abstraction of 
the $\sigma$-hole -- a three-dimensional object pertaining to 
halogen atoms. The statistical analysis of $\sigma$-hole magnitude, 
size, angular deviation $\phi$ and range has revealed a surprisingly 
low effect of intramolecular polarization on the spatial position 
of $V_{max}$. The other properties are much more affected. Our 
findings bring a novel, refined view on halogenated molecules and 
their ability to participate in noncovalent interactions and help
approach them since the analysed molecules represent a realistic
set for general use. Finally, we admit that because the $\sigma$-holes 
of atoms of IV, V and VI group are significantly more complicated 
in shape, similar analysis of them seems not to be as straightforward 
as for halogens.

MHK is grateful for support provided by the Alexander von Humboldt 
Foundation. This work was part of the Research Project RVO: 61388963
of the Institute of Organic Chemistry and Biochemistry, Academy of 
Sciences of the Czech Republic. It was also supported by the Czech
Science Foundation [P208/12/G016] and the operational program Research 
and Development for Innovations of the European Social Fund 
(CZ 1.05/2.1.00/03/0058).





\bibliography{ms.bbl}

\clearpage

\end{document}